\documentclass[10pt]{iopart}
\usepackage{iopams}
\usepackage{graphicx,amssymb,color,bm,hyperref}

\def\be{\begin{equation}}
\def\ee{\end{equation}}
\def\bea{\begin{eqnarray}}
\def\eea{\end{eqnarray}}
\newcommand{\vs}{\nonumber\\}
\def\ba#1\ea{\begin{eqnarray}#1\end{eqnarray}}

\newcommand{\g}{\gamma}
\newcommand{\refeq}[1]{Eq.~(\ref{eq:#1})}          
          
\newcommand{\refEq}[1]{Equation~(\ref{eq:#1})}          
          
\newcommand{\reffig}[1]{Fig.~\ref{fig:#1}}          
\newcommand{\refFig}[1]{Figure~\ref{fig:#1}}          
\newcommand{\refsec}[1]{\S~\ref{sec:#1}}

\renewcommand{\v}[1]{\mathbf{#1}}

\newcommand{\vx}{\v{x}}
\newcommand{\vr}{\v{r}}
\newcommand{\vk}{\v{k}}

\def\vtheta{\boldsymbol\theta}

\newcommand{\<}{\langle}
\renewcommand{\>}{\rangle}

\renewcommand{\k}{\kappa}
\renewcommand{\d}{\delta}
\newcommand{\D}{\Delta}

\newcommand{\nhat}{\hat{n}}
\newcommand{\vnhat}{\v{\hat{n}}}

\renewcommand{\l}{\ell}

\newcommand{\zt}{\tilde{z}}

\newcommand{\chib}{\bar{\chi}}
\newcommand{\chit}{\tilde{\chi}}

\def\Q{\mathcal{Q}}

\def\M{\mathcal{M}}
\def\P{\mathcal{P}}

\def\O{\mathcal{O}}

\def\A{\mathcal{A}}
\def\B{\mathcal{B}}
\def\C{\mathcal{C}}
\def\T{\mathcal{T}}

\def\Del{\eth}

\def\nhat{\hat{n}}
\def\vnhat{\hat{\v{n}}}

\definecolor{RedWine}{rgb}{0.743,0,0}
\definecolor{RoyalBlue}{rgb}{0.25,.41,.88}

\begin{document}

\review{Large-Scale Structure Observables in General Relativity}

\author{Donghui Jeong$^{1,2}$ and Fabian Schmidt$^3$}
\address{
${}^1$Department of Astronomy and Astrophysics,
The Pennsylvania State University, University Park, PA 16802, USA
}
\address{
${}^2$Institute for Gravitation and the Cosmos, The Pennsylvania State University, University Park, PA 16802, USA
}
\address{
${}^3$Max-Planck-Institut f\"ur Astrophysik, Karl-Schwarzschild-Str. 1, D-85741 Garching, Germany
}

\begin{abstract}
We review recent studies that rigorously define several key observables of the 
large-scale structure of the Universe in a general relativistic context. 
Specifically, we consider
i) redshift perturbation of cosmic clock events;
ii) distortion of cosmic rulers, including weak lensing shear and magnification;
iii) observed number density of tracers of the large-scale structure.  
We provide covariant and gauge-invariant expressions of these observables.  
Our expressions are given for a linearly perturbed 
flat Friedmann-Robertson-Walker metric including scalar, vector, 
and tensor metric perturbations.  
While we restrict ourselves to linear order in perturbation theory, 
the approach can be straightforwardly generalized to higher order.
\end{abstract}

\date{\today}

\submitto{\CQG}

\maketitle

\section{Introduction}
\label{sec:intro}

Over the past decade, cosmology has benefited from a vast increase
in the available data \cite{wmapfinal,wmapreview,planckoverview,
planckparameters,2dFBAO,sdssBAO}, 
which have been exploited through a variety of methods to probe the history 
and structure of the Universe. This trend will continue with future large 
surveys\footnote{
HETDEX ({\sf http://www.hetdex.org}), 
eBOSS ({\sf https://www.sdss3.org/future/eboss.php}), 
DESI ({\sf http://desi.lbl.gov}), 
SuMIRe ({\sf http://sumire.ipmu.jp/en/}), 
WFIRST-AFTA ({\sf http://wfirst.gsfc.nasa.gov}), 
Euclid ({\sf http://sci.esa.int/euclid/}), to list a few.}, 
and clearly, this calls for a rigorous investigation of what quantities 
precisely are observable from these surveys in the fully relativistic setting. 
Some observables have been investigated previously, most notably the 
number density of tracers \cite{yoo/etal:2009,yoo:2010,
bonvin/durrer:2011,challinor/lewis:2011,gaugePk,paperI}, the magnification, 
and to a lesser extent the shear \cite{DodelsonEtal,stdruler,paperII}.  

In this review, we present a unified relativistic analysis of the observables
from the large-scale structure of the Universe: \emph{standard clocks}, 
\emph{standard rulers}, and \emph{number density of galaxies} (more generally, tracers).  
First, we consider a clock comoving with the cosmic fluid that shows 
the proper time 
elapsed since the Big Bang.  The ``apparent'' age of the Universe at the same
location is inferred using the observed redshift of the source.  The difference
between the two is an observable \cite{Tpaper} that enters (often implicitly) 
in many different applications.  For example, the cosmic microwave background 
on large scales (Sachs-Wolfe limit) can be thought of as one 
of the cosmic clock events.

A standard ruler \cite{stdruler} simply means that there is
an underlying physical scale that we know of and we compare the observations 
to. This treatment applies to lensing measurements through galaxy ellipticities,
sizes and fluxes, or through standard candles, to distortions of 
cosmological correlation functions, and to lensing of diffuse backgrounds.   
We show that in this framework, for ideal measurements, one can
measure six degrees of freedom: a scalar on the sphere which corresponds to 
purely line-of-sight effects; a vector (on the sphere) which corresponds
to mixed transverse/line-of-sight effects; and a symmetric transverse
tensor on the sphere which comprises the shear and magnification \cite{paperII}.
We obtain general, gauge-invariant expressions for the six observable
degrees of freedom, valid on the full sky. Throughout this paper,
gauge invariance refers to the independence of the result on the choice
of global perturbed FRW coordinates 
(say, e.g., comoving gauge vs conformal-Newtonian gauge).  
That is, the expressions are directly applicable to compare with observations
and do not contain any unphysical coordinate artefacts.

The vector component and the shear admit a decomposition by parity into
$E/B$-modes.  The $B$-modes are free of all scalar contributions
(including lensing as well as redshift-space distortions) at the
linear level, making them ideal probes to look for tensor perturbations
(gravitational waves).  
Throughout, we will work to linear order in perturbations, although the
formalism can be straightforwardly extended to higher order.

Standard rulers in cosmology (again, think of mean size of a galaxy
sample, or the correlation length of a tracer), are rarely absolute constants.  
Rather, they evolve in time, and this time evolution
contributes at linear order to the observables described above through
the variation in proper time since the Big Bang on a constant observed
redshift surface.  If one observes two spatially co-located rulers that 
evolve differently in time, one can isolate this effect.  Thus,
standard rulers can serve as cosmic clocks in the sense discussed above as 
well. The contribution of this effect to the magnification that we include 
in the calculation is \emph{linear order} in metric perturbations but 
has previously been ignored.

Finally, we turn to another important large-scale structure observable,
the number density of tracers such as galaxies \cite{gaugePk,paperI}. 
This requires two ingredients:  
the transformation of the volume element from apparent
to physical volume, and the biasing relation between the density
contrast of tracers and matter perturbations in the chosen gauge.  The results
for standard rulers and clocks can immediately be applied to derive these two
contributions.

The outline of the paper is as follows: we begin in \refsec{geodesic} by
spelling out the expression for null geodesics in a perturbed FRW spacetime,
along with introducing our metric convention and useful notation.  
We then present the three large-scale observables in the subsequent sections:
cosmic clock (\refsec{clock}), cosmic ruler (\refsec{ruler}), and
galaxy number density (\refsec{clustering}).  We conclude in \refsec{conclusion}
with future perspective including the relevance of the effects to planned 
future surveys and further work needed to exploit the large-scale structure 
observables in general relativity.  
For all the quantitative results shown in this paper, we assume
a flat $\Lambda$CDM cosmology with $h=0.72$, $\Omega_m=0.28$, a scalar
spectral index $n_s=0.958$ and power spectrum normalization at $z=0$ of
$\sigma_8 = 0.8$.

\section{Light propagation in the perturbed universe}\label{sec:geodesic}

\subsection{Notation}\label{sec:not}

We use a conformal coordinate system $(\eta,x^i)$, and assume a spatially flat 
FRW background metric $\bar g_{\mu\nu} = a^2(\eta) \eta_{\mu\nu}$.  The
linearly perturbed FRW metric is written as
\be
\rmd s^2 = a^2(\eta)\left[
-(1+2A) \rmd\eta^2 - 2B_i \rmd\eta \rmd x^i 
+ \left(\d_{ij}+h_{ij}\right) \rmd x^i \rmd x^j \right]\,.
\label{eq:metric}
\ee
We will work to linear order in $A,\,B,\,h$ throughout.  
Latin indices $i,j,k,\cdots$ denote spatial coordinates, and Greek indices
$\mu,\nu,\rho,\sigma,\cdots$ denote the space-time coordinates.  Unless
otherwise indicated, we raise and lower space-time indices with the full 
metric \refEq{metric}, and space indices by $\delta_{ij}$.  For scalar
perturbations, the spatial part is often further expanded as
$h_{ij} = 2(D \d_{ij} + E_{ij})$, where $E_{ij}$ is symmetric and traceless.  
We denote the trace of $h_{ij}$ as $h \equiv \d^{ij} h_{ij}$.  
In \refsec{ruler}, we shall also present results in two popular
gauges: the synchronous-comoving (sc) gauge, where $A = 0 = B_i$, so that
\be
ds^2 = a^2(\eta)\left[- d\eta^2
+ \left(\d_{ij}+h_{ij}\right) dx^i dx^j \right];
\label{eq:metric_sc}
\ee
and the conformal-Newtonian (cN) gauge, where $B_i = 0 = E_{ij}$, so that
(with $A = \Psi$, $D = \Phi$)
\be
ds^2 = a^2(\eta)\left[- (1+2\Psi) d\eta^2
+ (1+2\Phi) \d_{ij} dx^i dx^j \right].
\label{eq:metric_cN}
\ee

It is useful to define projection operators parallel ($\parallel$) and 
perpendicular ($\perp$)
to the observed line-of-sight direction $\nhat^i$,
so that for any spatial vector $X^i$ and tensor $h_{ij}$,
\ba
X_\parallel \equiv\:& \nhat_i X^i, ~
h_\parallel \equiv\: \nhat_i \nhat_j h^{ij}, \vs
X_\perp^i \equiv\:& \P^{ij} X_j , ~
\P^{ij} \equiv\: \d^{ij} - \nhat^i \nhat^j.
\label{eq:proj1}
\ea
Correspondingly, we define projected derivative operators,
\be
\partial_\parallel \equiv\: \nhat^i\partial_i,{\rm ~and~} 
\partial_\perp^i \equiv\: \P^{ij} \partial_j.
\label{eq:proj2}
\ee
Note that $\partial_\perp^i,\,\partial_\parallel$ and $\partial_\perp^i,\,\partial_\perp^j$
do not commute, while $\nhat^i$ and $\partial_\parallel$ do commute.  
Further, we find
\be
\partial_j \nhat^i = \partial_{\perp j} \nhat^i = \frac1\chi \P_j^{\  i},
\ee
where $\chi$ is the norm of the position vector so that $\nhat^i = x^i/\chi$.  
More expressions can be found in \S~II of \cite{gaugePk}.

We further decompose quantities defined on the
sphere, i.e. functions of the unit line-of-sight vector $\vnhat$, 
in terms of their properties under a rotation around $\vnhat$.  Let $(\v{e}_1,\v{e}_2,\vnhat)$ denote an orthonormal coordinate system.  
If we rotate the coordinate system around $\vnhat$ by an angle $\psi$,
so that $\v{e}_i \to \v{e}'_i$,
then the linear combinations $\v{m}_\pm \equiv (\v{e}_1\mp i\,\v{e}_2)/\sqrt{2}$
transform as 
\be
\v{m}_\pm \to \v{m}'_\pm = e^{\pm i\psi} \v{m}_\pm\,.
\label{eq:spin1}
\ee
A general function, or, more properly, tensor component $f(\vnhat)$ 
is called spin-$s$ if it transforms under the same transformation as
\be
f(\vnhat) \to f(\vnhat)' = e^{i s\psi} f(\vnhat)\,.
\ee
An ordinary scalar function on the sphere is clearly spin 0, while the
unit vectors $\v{m}_\pm$ defined above are spin$\pm1$ fields.
This decomposition is particularly
useful for deriving multipole coefficients and angular power spectra.  
We also define
\be
X_\pm \equiv\: m_\mp^i X_i,~
h_\pm \equiv\: m_\mp^i m_\mp^j h_{ij}
\label{eq:Xpm}
\ee
for any 3-vector $X_i$ and 3-tensor $h_{ij}$.

\subsection{Integration of geodesic equation}
\label{sec:deflection}

\begin{figure}[t!]
\centering
\includegraphics[width=0.45\textwidth]{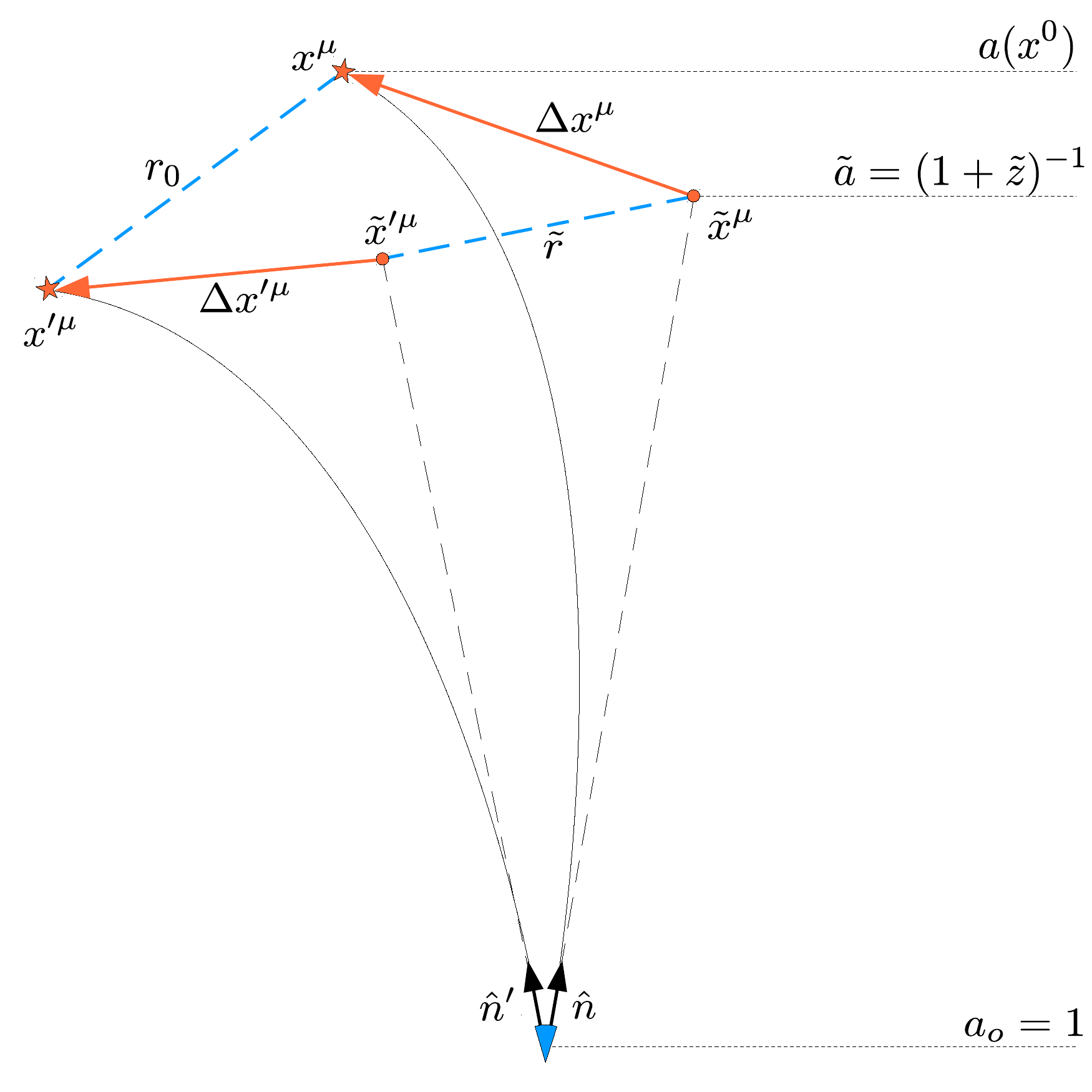}
\caption{
Sketch of perturbed photon geodesics illustrating our notation
(from \cite{stdruler}).  
The observer is located at the bottom.  Photons
arrive out of the observed directions $\vnhat,\,\vnhat'$ and with
observed redshifts $\zt,\,\zt'$.  The solid lines indicate the
actual photon geodesics tracing back to the sources indicated by stars.  
The dashed lines show the apparent background photon geodesics tracing back to
the inferred source positions indicated by circles, perturbed from the
actual positions by the displacements $\D x^\mu,\,\D x'^\mu$
(whose magnitude is greatly exaggerated here).  
$\tilde r$ is the apparent spatial separation between the endpoints
(apparent size of the ruler), while $r_0$ is the true separation.  
\label{fig:sketch}}
\end{figure}

Cosmological observations are made by collecting light from distant sources
(e.g. galaxies) and measuring their positions, fluxes, shapes, and redshifts.  
The relativistic effects we discuss in this paper are due to 
the displacement between the intrinsic spacetime location of sources and their 
apparent location inferred from the observed angular coordinate 
$\vnhat$ and redshift $z$ by using the background FRW metric (see \reffig{sketch}). 
In this section, we outline the derivation of the displacements for a 
source comoving with the cosmic fluid\footnote{The assumptions of comoving source and observer are easily relaxed to allow for general motion.} (see also \cite{PyneBirkinshaw}) described by the four velocity
\be
u^\mu = a^{-1} \left(1-A,v^i\right),~
u_\mu = a\left(-1-A,v_i-B_i\right)
\label{eq:umu}
\ee
in the general gauge given by \refEq{metric}.  
Note that we take care to include all the contributions explicitly, including 
those which only contribute to the monopole.  These contributions are essential
when verifying the expressions through test cases as described in the 
appendices of \cite{gaugePk,stdruler,Tpaper}.  

Choosing the background comoving distance $\chi$
as affine parameter, the conformal photon momentum may be written as
\be
\frac{dx^\mu}{d\chi} = \big(-1+\d\nu(\chi),\: \nhat^i + \d e^i(\chi)\big),
\label{eq:phmom}
\ee
with the fractional frequency shift $\delta\nu$ and the deflection $\delta e^i$
of the photon along the geodesic.
The light propagation in the perturbed FRW metric is described by 
the geodesic equation 
\ba
\frac{d}{d\chi}\frac{dx^\mu}{d\chi} + \Gamma^\mu_{\alpha\beta} 
\frac{dx^\alpha}{d\chi}\frac{dx^\beta}{d\chi} =\:0.
\ea

The zeroth order geodesic equation is just $dx^\mu/d\chi={\rm constant}$, 
which yields the background geodesic,
\be
\bar{x}^\mu(\chi)
=
\left(
\eta_0-\chi, \vnhat\chi
\right),
\label{eq:geod_conf}
\ee
where $\eta_0$ is the conformal time at present where the observation is made.
\refEq{geod_conf} determines the apparent position 
$\tilde{x}^\mu$ of a source with observed redshift $\zt$ and angular 
position $\nhat^i$ as 
$\tilde{x}^\mu =\: (\eta_0 - \chit, \nhat^i\,\chit)$ 
with 
$\chit \equiv\: \chib(\zt)$
being the comoving distance-redshift relation in the
background Universe. 
Hereafter, we shall use tilde to refer to the observed (or apparent)
quantities, and bar to refer to background quantities.

At linear order, the temporal and spatial components of the geodesic equation 
read
\ba
\fl
\frac{d}{d\chi} \left(\d\nu - 2 A\right) = \dot{A} 
- \frac12 \dot{h}_{\parallel} - \partial_\parallel B_\parallel
\label{eq:geodesicT}
\\
\fl
\frac{d}{d\chi} \left(\d e^i + B^i + h^i_{\  j}\nhat^j \right) =
- \partial_i A + \partial_i B_\parallel - B_{\perp i} 
+ \frac12 \partial_ih_\parallel - \frac1\chi \P^{ij} h_{jk} \nhat^k.
\label{eq:geodesicS}
\ea
Here and throughout, a dot represents the derivative with respect to
conformal time.  
We integrate the geodesic equation starting from the comoving observer 
at $\chi=0$. We choose him or her to lie at the spatial 
origin $\v{0}$, and to observe at a fixed \emph{proper time} 
$t_o$ [see \refeq{tF} in the next section]:
\be
t_o = \int_0^{\eta_o} \left[1+A(\v{0},\eta')\right] a(\eta')d\eta'.
\label{eq:t0}
\ee
Note that this is a coordinate independent way of defining the observation 
time, and we normalize the scale factor with the proper time at observation 
through $a(t_o)=1$ following the usual convention.  Then, the scale factor 
in the metric \refEq{metric} at observation differs from the 
background value by $\d a_o = a_o -1$:
\be
\d a_o = - \frac{da}{dt}\Big|_o \int_0^{t_o} A(\v{0},\bar\eta(t)) dt
= - H_0 \int_0^{t_o} A(\v{0},\bar\eta(t)) dt\,.
\label{eq:delta_a_o}
\ee
In what follows, the subscript ${}_o$ will refer to quantities evaluated at 
the observer's spacetime location. 

The initial conditions of $\d\nu$, $\d e^i$ at the observation are fixed by 
the requirement that the observed (past-directed) photon four momentum is 
$(1,\nhat^i)$.  
An orthonormal tetrad $(e_a)^\mu$ for an observer comoving with the cosmic 
fluid (as we will assume throughout) is given by
\ba
(e_0)^\mu =\:& a^{-1} \left(1-A,\, v^i \right) \vs
(e_j)^\mu =\:& a^{-1} \left(v_j-B_j,\, \d_j^{\  i} - \frac12 h_j^{\  i}\right).
\label{eq:tetrad}
\ea
Here, we use $a,b=0,1,2,3$ for the space-time index of the tetrad
$(e_a)$, while $\mu,\nu=0,1,2,3$ denotes the coordinate index of the tetrad $(e_a)^\mu$. Then, the observed energy and momentum at $\chi=0$ are, respectively,
\be
1 =\: \left[a^{-2}\, g_{\mu\nu}(e_0)^\mu\: p^\nu\right]_o,~
\nhat_i =\: \left[a^{-2}\, g_{\mu\nu}(e_i)^\mu p^\nu\right]_o,
\label{eq:pobs}
\ee
from which we find the initial conditions as
\ba
\d\nu_0 =\:& -\d a_o + A_o + v_{\parallel o} - B_{\parallel o} \vs
\d e^i_o =\:& \d a_o \nhat^i - v^i_o - \frac12 (h^i_{\  j})_o \nhat^j.
\label{eq:geodesicIC}
\ea
The $a^{-2}$ factors in \refeq{pobs} come from the transformation of the affine 
parameter with respect to the conformal metric ($\chi$) to that corresponding 
to the physical metric ($\lambda$) through 
$d\chi/d\lambda = a^{-2}$ \cite{gaugePk}.  

Given the initial conditions \refEq{geodesicIC}, we 
now integrate the geodesic equation twice to yield the 
temporal and spatial displacements:
\ba
\fl
\d x^0(\chi)
=& \left[-\d a_o - A_o + v_{\parallel o} - B_{\parallel o}\right]\chi
\label{eq:dx0G}\\
\fl
&+ \int_0^{\chi} d\chi'\left[ 2 A + (\chi-\chi') \left\{
\dot{A} - \frac12 \dot{h}_{\parallel} - \partial_\parallel B_\parallel\right\}
\right] 
 - \int_0^{t_o} A(\v{0}, t) dt 
\; \vs
\fl
\d x^i(\chi) 
=& \left[\d a_o \nhat^i + \frac12 (h^i_{\  j})_o\, \nhat^j + B^i_o - v^i_o\right] \chi
\label{eq:dxiG}\\
\fl
&+ \int_0^{\chi} d\chi' \left[
- B^i - h^i_{\  j}\nhat^j + (\chi-\chi')\left\{
-\partial_i A +\nhat^j\partial_iB_{j}
+ \frac12 (\partial_i h_{jk}) \nhat^j\nhat^k
\right\}\right].
\nonumber
\ea
Here, we used $\delta x^i(\chi=0)=0$ as spatial boundary condition at
the observer. The temporal boundary condition employed in \refeq{dx0G} 
is determined through the fixed proper time of observation, \refEq{t0},
and  
$\d x^0(\chi=0) = \eta_o - \bar\eta(t_o)$ where $\bar{\eta}(\bar{t})$ is the 
background conformal time--physical time relation.  

The final ingredient that we need to complete the calculation is 
the affine parameter at emission of the photon, $\chi_e = \chit + \d\chi$. 
This is fixed by requiring the photon frequency to match the observed redshift $\zt$.  
From the tetrad in \refEq{tetrad}, $\zt$ is given by 
the ratio between the photon energy at emission and observation as:
\be
1 + \zt \equiv \frac{1}{\tilde{a}}
=
\frac{\left[
a^{-2} g_{\mu\nu}\left(e_o\right)^\mu p^\nu
\right]_e}
{\left[
a^{-2} g_{\mu\nu}\left(e_o\right)^\mu p^\nu
\right]_o}
= \frac{1}{a(x^0)} (1 + A - \d\nu + v_\parallel - B_\parallel).
\qquad
\label{eq:zmatch}
\ee
Here, all quantities on the right hand side are evaluated at emission.  
We find it convenient to define the perturbation to the logarithm of the 
scale factor at emission $\Delta \ln a \equiv a(x^0)/\tilde{a}-1$.  
\refEq{zmatch} yields
\be
\fl
\D\ln a 
=\: A_o - A + v_\parallel - v_{\parallel o} 
+ \int_0^{\chit} d\chi\left[
- \dot{A} + \frac12 \dot{h}_{\parallel} + \dot{B}_\parallel\right]
- H_0 \int_0^{t_o} A(\v{0},\bar\eta(t)) dt\,.
\label{eq:Dlna2}
\ee
The redshift matching then requires 
\be
\Delta \ln a 
=\left.\frac{\partial \ln a}{\partial \eta}\right|_{\zt} (x^0 - \tilde x^0)
= \frac{H(\zt)}{1+\zt} (\delta x^0(\chit) - \delta\chi)\,.
\ee
We can then solve for the perturbation to the affine parameter yielding
\be
\delta \chi = \delta x^0(\chit) - \frac{1+\zt}{H(\zt)}\Delta\ln a.
\ee
The actual source position is at linear order
\ba
\fl
x^\mu(\chi_e) 
= \bar{x}^\mu(\chi_e) + \delta x^\mu(\chit)
= \bar{x}^\mu(\chit) 
+ \left[\bar{x}^\mu(\chi_e) - \bar{x}^\mu(\chit)\right]
+ \delta x^\mu(\chit).
\ea
To linear order, this gives
\ba
\Delta x^0 \equiv&
x^0 - \tilde{x}^0 = \delta x^0(\chit) -\delta\chi
\vs
\Delta x^i \equiv&
x^i - \tilde{x}^i = \delta x^i(\chit) +\nhat^i\delta\chi.
\ea
Finally, we assemble the deflection along the line of sight direction as
\ba
\fl
\D x_\parallel &= \nhat_i\Delta x^i 
= \d x_\parallel + \d x^0 - \frac{1+\zt}{H(\zt)}\D\ln a
\vs
\fl
&= -\int_0^{t_o} A(\v{0},t) dt + \int_0^{\chit} d\chi\left[ A - B_\parallel - \frac12 h_\parallel
\right]
- \frac{1+\zt}{H(\zt)} \D \ln a\,.
\label{eq:DxparG}
\ea
The transverse displacement $\D x_\perp^i = \P^i_{\  j}\Delta x^j$ is given by
\ba
\fl
\D x_\perp^i  =& \left[\frac12 \P^{ij} (h_{jk})_o\, \nhat^k + B^i_{\perp o} - v^i_{\perp o}\right] \chit \label{eq:Dxperp2}
- \int_0^{\chit} d\chi \bigg[
\frac{\chit}{\chi} \left(
B_\perp^i + \P^{ij} h_{jk}\nhat^k\right) 
\vs
\fl
&+(\chit-\chi)\partial_\perp^i
\left( A - B_\parallel - \frac12 h_\parallel\right)
\bigg].
\label{eq:DxperpG} 
\ea

\section{Cosmic clock}\label{sec:clock}

Perhaps the simplest general relativistic observables are cosmic clocks \cite{Tpaper}, 
or standard clocks, a set of events whose proper time 
(local age of the universe) is accurately known.
These cosmic clock events define an observable \emph{$\T(\nhat)$ as the 
difference in logarithm of scale factor $(\ln a)$ between 
a constant-proper-time hypersurface $t_F={\rm const}$ and a 
constant-observed-redshift surface $\tilde{z}={\rm const}$}.
Although phrased as a perturbation in $\ln a$, we will frequently refer to $\T$
loosely as the proper time perturbation. This is because at leading order, 
the perturbation to the proper time $\D t_F(\vnhat)$ at observed redshift
$\zt$ is simply related to $\T$ through
\be
\D t_F(\vnhat) = H^{-1}(\zt) \T(\vnhat)\,.
\ee
Since $\T$ is defined through two observationally well-defined 
quantities (proper time and observed redshift), 
it is  an observable, and the expression for $\T$ is gauge-invariant.

The proper time interval $dt_F$ along the geodesic of a comoving
observer is at linear order given by
\be
dt_F = \sqrt{- g_{\mu\nu} dx^\mu dx^\nu} = (1 + A) a d\eta\,.
\label{eq:dtF}
\ee
Integrating \refEq{dtF},
we obtain an expression for $t_F|_x$, the proper time of a
comoving source passing through $\vx$ at coordinate time $x^0$,
at linear order 
\ba
t_F|_x =\:& \int_0^{x^0} \left[1 + A(\vx,\eta') \right] a(\eta') d\eta' \,.
\label{eq:tF}
\ea
In the case at hand, $x^0$ is the coordinate time at emission, which 
is different from the coordinate time $\bar{\eta}(t_F|_x)$ 
in an unperturbed universe at the proper time $t_F$.  
The ratio of scale factors at coordinate time $x^0$ and $\bar{\eta}(t_F|_x)$ is
\be
\frac{a\left[\bar\eta(t_F|_x)\right]}{a(x^0)} 
=\: 1 + H(x^0) \int_0^{x^0} A(\vx,\eta') a(\eta') d\eta'\,.
\label{eq:aratio}
\ee
The observable $\T(\tilde x)$ is defined through
\be
\T(\tilde x) \equiv \ln \left(\frac{a\left[\bar\eta(t_F|_{\tilde x})\right]}{\tilde a}\right) \,,
\ee
where $\tilde a = (1+\zt)^{-1}$ is the apparent scale factor at emission.  
Since $\ln[ a(x^0)/\tilde a]$ is precisely the perturbation $\D\ln a$ derived in \refsec{geodesic}, 
we arrive at the following simple expression for $\T$:
\ba
\T =\:&
\ln \left(\frac{a(x^0)}{\tilde a} 
\frac{a\left[\bar\eta(t_F|_{\tilde x})\right]}{a(x^0)}
\right)
= \D\ln a +\tilde H \int_0^{\tilde\eta} A[\vx,\eta'] a(\eta') d\eta' 
\,,
\label{eq:Tdef}
\ea
where $\tilde H \equiv H(\zt)$.  
The gauge invariance of the observable $\T$ is explicitly shown in 
the Appendix A of \cite{Tpaper}.  The numerical result for the
power spectrum of $\T$ induced by a standard power spectrum of scale-invariant
curvature perturbations in $\Lambda$CDM is shown in \reffig{Cl}.  

A cosmic clock exists whenever we have an observable from which we can define 
a constant proper-time hypersurface.
There are two such classes of observables: 1) cosmic events defined by
a unique time and sufficiently short duration, 2) observables with known
time evolution.  
Cosmic recombination, which led to the emission of the cosmic microwave 
background (CMB) is an example of the former.  Neglecting all perturbations
including sound waves, last scattering of the Cosmic Microwave Background (CMB) photons
happened at a fixed proper time in the local frame, and, therefore, is a
cosmic clock event.  On scales larger than the angular size of the sound 
horizon at recombination, the temperature perturbations in the CMB are
then exactly given by $\Theta(\nhat) = - \T(\nhat)$ (Sachs-Wolfe limit \cite{SW67,Tpaper}).  
The exact same equation applies in describing the perturbations 
in, for example, the surface of neutrino decoupling, 
Big-Bang Nucleosynthesis and thermal decoupling of baryons from CMB 
on super-horizon scales.  
The second case is exemplified by time varying cosmic rulers, 
which will be discussed in the next section.

\section{Cosmic ruler, or generalized weak gravitational lensing}
\label{sec:ruler}

We now move on to the cosmic ruler, with which we mean a known 
spatial scale in the comoving frame of the cosmic fluid.  
This could be the size of a galaxy, or 
the length at which the correlation function reaches a certain value  
(we will discuss these applications in \refsec{app}).  
What we observe is the \emph{apparent} size of this 
``ruler'', inferred using the observed positions and redshifts of the 
endpoints.  
By comparing this to the known spatial scale, we can infer the departure 
from the average angular diameter distance-redshift relation and
the Hubble parameter-redshift relation.
In many cases, the size of the ruler will only be known in a statistical
sense (for example, galaxy sizes) and will be calibrated by averaging
over the entire area of a given survey.  Any scatter in the ruler size
from its mean value will simply be noise in the measurement of the
ruler distortions, as long
as this scatter is not correlated with large-scale perturbations themselves.  
The latter, ``intrinsic'' contributions to the distortion of the ruler
scale will not be considered in this paper, although they can be an
important source of cosmological information on their own 
\cite{CatelanKamionkowskiBlandford,HirataSeljak04,BrownEtal02,BlazekEtal,JoachimiEtal,pen/etal:12,paperII,schmidt/pajer/zaldarriaga}.

As we shall show below, the ruler distortions can be decomposed into 
scalar, vector, and tensor components on the observer's sky.  
The tranverse ``sky-plane'' components, one scalar and two tensor 
components, are nothing other than the standard lensing observables of 
magnification and shear, respectively.  
The remaining distortions, one scalar and two vector components on the sky,
involve the line-of-sight component of the ruler.  
Typically, these can only be observed through spectroscopic
measurements, since the line-of-sight separation, inferred from the
redshift difference between the two endpoints, has to be 
measured with sufficient accuracy.  
When applied to correlation functions, these distortions are part of the 
well-known redshift-space distortion effects.  However,
we stress that the expressions we derive are entirely independent of the
nature of the ruler considered, and that spectroscopic LSS surveys are just
one (albeit important) application of these new observables.

A standard ruler can be generically modeled by two spacetime events 
separated by a fixed spacelike separation $r_0$ on a fixed proper time
surface of the cosmic fluid.  More precisely, the spatial part of
the four-velocity $u^\mu$ of this fluid is determined by
\be
v^i = \frac{T^i_{\  0}}{\rho + p}\,.  
\label{eq:vcom}
\ee  
This ruler definition can also be phrased as that the length of the ruler 
is defined on a surface of constant proper time of comoving observers.  
This proper time corresponds to the ``local age'' of the Universe.  
We are mostly interested in applications to the large-scale structure during
matter domination.  In this case, the cosmic fluid is simply matter
(dark matter + baryons), and there is no ambiguity in this definition;  
in synchronous-comoving gauge, \refEq{vcom} yields $v^i = 0$.  
However, this assumption can be relaxed very easily, for example one could
assume instead that the observers are comoving with the baryon velocity $v_b$.  

Then, what we observe is the apparent size of the ruler.  
Let $\vnhat,\tilde z$ and $\vnhat',\tilde z'$ denote the observed 
coordinates of the endpoints of the ruler, and $\tilde\vx$ and
$\tilde\vx'$ the apparent spatial positions inferred through \refEq{geod_conf} (see \reffig{sketch}).  
In the following, we will assume that the ruler is small compared to the
distance $\chit$ of the sources as well as to the
typical scale over which we want to measure the spacetime perturbations; 
it can then be approximated as an infinitesimal distance.  
For example, in terms of weak lensing observables, we assume that the angular size of a 
galaxy is negligible compared to the angular scale at which we measure
shear correlations.  
Corrections to this approximation involve powers of $|\tilde\vx-\tilde\vx'|/\chit$ (wide-angle corrections), and/or higher derivatives
of the metric perturbations multiplied by powers of $x^\mu - x'^\mu$.\footnote{For example, for the sky-plane components the leading order correction of this type corresponds to the lensing flexion.  If desired one could straightforwardly extend the treatment to obtain a covariant expression for the flexion.}
The apparent physical length of the cosmic ruler is then given by
\be
\tilde r^2 = \tilde a^2 \d_{ij} (\tilde x^i - \tilde x'^i) (\tilde x^j - \tilde x'^j)\,,
\label{eq:rt1}
\ee
where $\tilde a = 1/(1+\zt)$ is the observationally inferred scale factor
at emission (\reffig{sketch}).  The actual separation of the two endpoints
of the ruler, $x^\mu,\,x'^\mu$, as measured in the comoving frame, 
on the other hand should be equal to the fixed scale $r_0$:
\ba
\left[g_{\mu\nu}(x^\alpha) + u_\mu(x^\alpha) u_\nu(x^\alpha)\right]
 (x^\mu-x'^\mu) (x^\nu - x'^\nu) = r_0^2\,.
\label{eq:r01}
\ea
The four-velocity of comoving observers, whose spatial components
are fixed by \refEq{vcom}, is given by \refeq{umu}.  
With this, \refEq{r01} yields
\ba
 -2 \tilde a^2 v_i \left\{ \d\tilde x^0 \d\tilde x^i + \d\tilde x^0 [\D x^i -\D x'^i] + \d\tilde x^i [\D x^0 - \D x'^0 ] \right\} &\vs
+ g_{ij}(x^\alpha) \bigg\{ \d\tilde x^i \d\tilde x^j + \d\tilde x^i [ \D x^j - \D x'^j ] 
+ [ \D x^i - \D x'^i ]\d\tilde x^j
\bigg\}
& = r_0^2,
\label{eq:r02}
\ea
where $\D x^\mu = \D x^\mu(\vnhat,\zt)$, $\D x'^\mu = \D x^\mu(\vnhat',\zt')$,
and the components of the \emph{apparent} separation vector are
$\d\tilde x^\mu = \tilde x^\mu - \tilde x'^\mu$.  
In order to evaluate the spatial metric $g_{ij}(x^\alpha)$ at the 
location of the ruler, we use $\Delta \ln a = a(x^0)/\tilde{a}-1$
to obtain at first order
\be
g_{ij}(x^\alpha) = \tilde a^2\left[\left(1 + 2 \D\ln a \right) \d_{ij} + h_{ij} \right].
\ee
We now again make use of the ``small ruler'' approximation, so that
\be
\D x^i - \D x'^i \simeq \d\tilde x^\alpha \frac{\partial}{\partial\tilde x^\alpha}
\D x^i.
\ee
Like any vector, we can decompose the spatial part of the apparent 
separation $\d\tilde x^i$ into parts parallel and transverse to the line
of sight:
\ba
\d\tilde x_\parallel \equiv\:& \nhat_i \d\tilde x^i \vs
\d\tilde x_\perp^i \equiv\:& \P^i_{\  j} \d\tilde x^j = \d\tilde x^i - \nhat^i \d\tilde x _\parallel.
\ea
In the correlation function literature, $\d\tilde x_\parallel,\,|\d\tilde\vx_\perp|$
are often referred to as $\pi$ and $\sigma$, respectively.  Then,
\ba
\d\tilde x^\alpha \frac{\partial}{\partial\tilde x^\alpha} =\:&
 (\d\tilde x^0 \partial_\eta + \d\tilde x_\parallel \partial_\parallel) 
+ \d\tilde x_\perp^i \partial_{\perp\,i},
\ea
where we have similarly defined $\partial_\parallel = \nhat^i \partial_i$,
$\partial_{\perp\,i} = \P^{\  j}_i \partial_j$.   Since the observed
coordinates $\tilde x^\mu$ by definition satisfy the light cone condition 
with respect to the unperturbed FRW metric, 
we have $\d\tilde x^0 = -\d\tilde x_\parallel$ in the small-angle approximation.  
Thus,
\ba
\d\tilde x^0 \partial_\eta + \d\tilde x_\parallel \partial_\parallel
=\:& \d\tilde x_\parallel (\partial_\parallel - \partial_\eta) 
=\:& \d\tilde x_\parallel \frac{\partial}{\partial\chit} = \d\tilde x_\parallel
H(\zt) \frac{\partial}{\partial\zt},
\ea
where $\partial/\partial\chit$ is the derivative with respect to the affine
parameter at emission.  We thus have
\be
\d\tilde x^\alpha \frac{\partial}{\partial\tilde x^\alpha} = 
\d\tilde x_\parallel \partial_{\chit} + \d\tilde x_\perp^i \partial_{\perp\,i}.
\ee
Working to first order in perturbations, we then obtain 
\ba
r_0^2 - \tilde r^2  =\:& 2 \D\ln a\: \tilde r^2 
+ \tilde a^2  h_{ij} \d\tilde x^i \d\tilde x^j 
+ 2 \tilde a^2 \left( v_\parallel  \d\tilde x_\parallel^2 + v_{\perp\,i} \d\tilde x_\perp^i \d\tilde x_\parallel\right)
\vs
&+ 2\tilde a^2 \d_{ij} \d\tilde x^{i} \left(\d\tilde x_\parallel \partial_{\chit}
+ \d\tilde x_\perp^k \partial_{\perp\,k} \right) \D x^{j}.
\label{eq:rt2}
\ea
All terms are straightforward to interpret: there are the perturbations
to the metric at the ruler location (both from the metric perturbation $h_{ij}$ and the
perturbation to the scale factor at emission); the contribution $\propto v$
from the projection from fixed-$\eta$ to fixed-proper-time
hypersurfaces; and the difference in
the spatial displacements of the endpoints of the ruler.  

\subsection{Clocks and evolving rulers}
\label{sec:evolving_ruler}
In the previous section, we have implicitly assumed, as is usually done,
that the ruler scale is constant in time, i.e. a non-evolving ruler.  
However, in many instances in cosmology, rulers do evolve over time; 
that is, the ruler scale $r_0$ depends on the local age of the Universe 
(proper time of the comoving observer).
For example, the mean size of galaxies evolves, and so does the correlation
length of large-scale structure tracers.  Thus, cosmic rulers in general are
also cosmic clocks (\refsec{clock}).  

Let us consider a standard ruler whose length evolves in time. Then,
by using \refEq{aratio}, we can parametrize the time evolution of the proper
size of the standard ruler $r_0(a)$ 
through its value in an unperturbed Universe as function of the scale 
factor $a$. The actual proper size of the ruler $r_0(a(t_F|_{x}))$, 
which is the size of the ruler in the constant-proper-time slicing,
relative to the size it evaluates to when inserting the apparent scale factor
of emission $\tilde a = (1+\zt)^{-1}$, in the constant-observed-redshift
slicing, is given by
\ba
\frac{r_0(a\left(t_F|_{x}\right))}{r_0(\tilde a)} 
= 1 + \frac{d\ln r_0(\tilde a)}{d\ln \tilde a} \T(\tilde x) \,,
\label{eq:r0evolv}
\ea
where $\T$ is defined in \refeq{Tdef}.
Note that we are assuming that $a_o = 1$ at observation, \refEq{delta_a_o},
so that $r_0(1)$ 
corresponds to the ruler scale today as calibrated by the observer.  
This clearly requires that $\T=0$ for a locally measured ruler, which is
the case for \refeq{Tdef}.  To reach this consistency, it is essential that
the epoch of observation $t_o$ is fixed in terms of proper time, rather than 
coordinate time, as discussed in \refsec{geodesic}.   

We thus have an additional contribution to the ruler distortion \refEq{rt2}
which is proportional to the time derivative of the ruler, explicitly
\be
- 2 \T \frac{d\ln r_0(\tilde a)}{d\ln\tilde a}\tilde r^2\,.
\ee

\subsection{Scalar-vector-tensor decomposition on the sky}
\label{sec:SVT}

\begin{figure}
\centering
\includegraphics[width=0.5\textwidth]{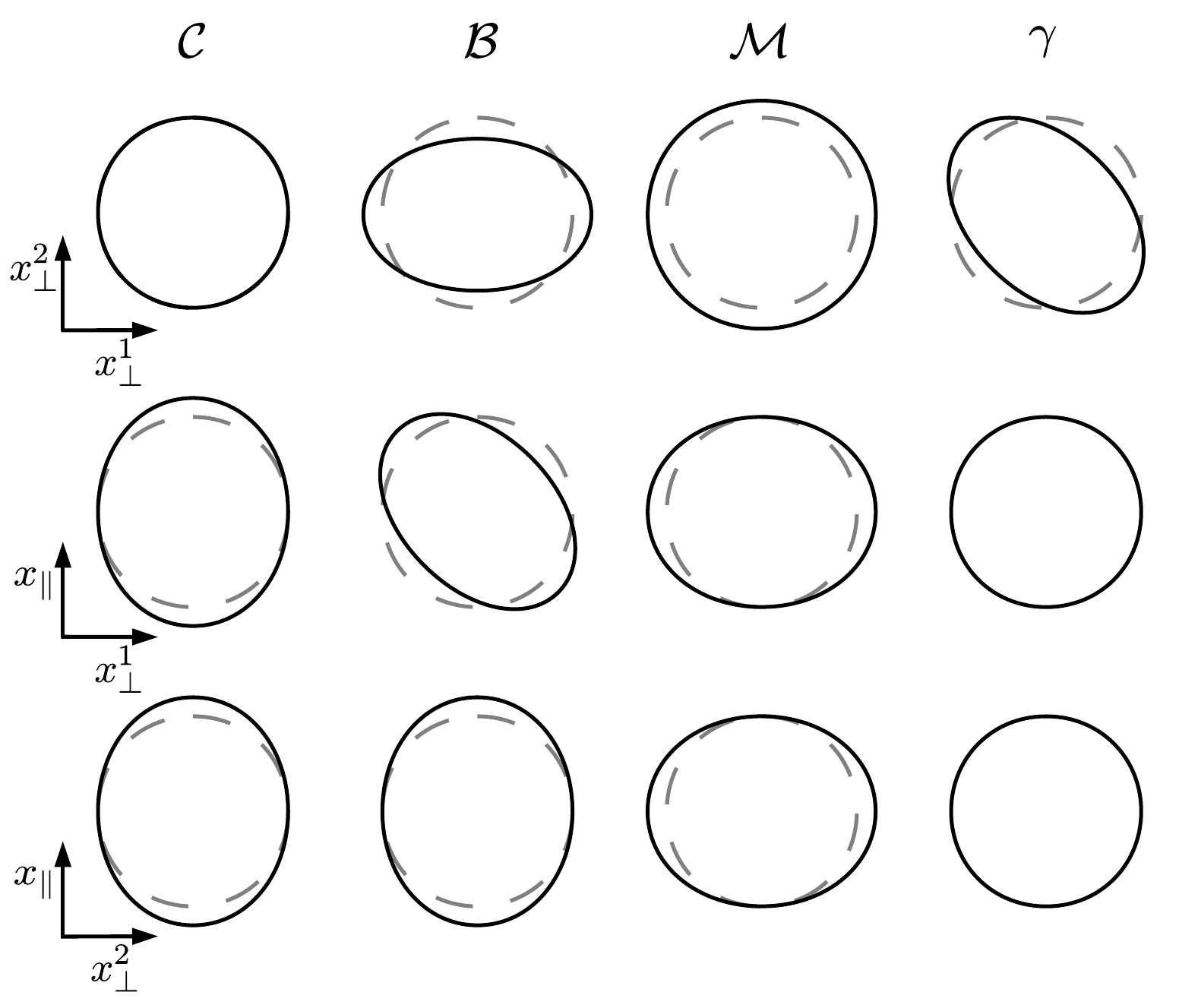}
\caption{Illustration of the distortion of standard rulers
due to the longitudinal (2-)scalar $\mathcal{C}$, (2-)vector $\B$, and 
transverse components, magnification $\M$ and shear $\g$.  
The first row shows the projection onto the sky plane, while the
second (third) row show the projection onto the line-of-sight and
$x_\perp^1$ ($x_\perp^2$) axes, respectively.  In case of $\B$ and $\g$,
we only show one of the two components.  From \cite{stdruler}; see also Fig.~3 in \cite{sachs:1961}.  
\label{fig:SVT}}
\end{figure}

It is useful to separate the contributions to \refEq{rt2} in terms of the 
observed longitudinal and transverse distortions.  For some
applications, only the transverse distortions are relevant.  This
is the case for diffuse backgrounds without redshift resolution,
such as the CMB or the cosmic infrared background, and largely the case
for photometric galaxy surveys.  On the other hand, spectroscopic
surveys and redshift-resolved backgrounds such as the 21cm emission
from high-redshifts are able to measure the longitudinal displacements
as well.  

Noting that
$\tilde r^2 = \tilde a^2[\d\tilde x_\parallel^2 + (\d\tilde\vx_\perp)^2]$, 
and taking the square root of \refEq{rt2}, we obtain the 
relative perturbation to the physical scale of the ruler as
\ba
\frac{\tilde r - r_0 }{\tilde r} =\:&
\C \frac{(\d\tilde x_\parallel)^2}{\tilde r_c^2} + \B_i \frac{\d\tilde x_\parallel \d\tilde x_\perp^i}{\tilde r_c^2}
+ \A_{ij} \frac{\d\tilde x_\perp^i \d\tilde x_\perp^j}{\tilde r_c^2},
\label{eq:r03}
\ea
where we have defined
$\tilde r_c \equiv \tilde r/\tilde a$ as the apparent comoving size of the ruler. 
The quantities multiplying $\C,\,\B_i,\,\A_{ij}$ are thus simply geometric
factors.  The coefficients are given by
\ba
\C =\:&  \frac{d\ln r_0(\tilde a)}{d\ln\tilde a} \T - \D\ln a 
- \frac12 h_\parallel - v_\parallel - \partial_{\chit} \D x_\parallel
\vs
\B_i =\:& -\P_i^{\  j} h_{jk} \nhat^k - v_{\perp i} - \nhat^k \partial_{\perp\,i} \D x_k - \partial_{\chit} \D x_{\perp i}
\vs
\A_{ij} =\:& \frac{d\ln r_0(\tilde a)}{d\ln\tilde a} \T\,\P_{ij}
- \D\ln a\: \P_{ij}
- \frac12 \P_i^{\  k}\P_j^{\  l} h_{kl} \vs
& - \frac12 \left(\P_{jk} \partial_{\perp\,i} + \P_{ik} \partial_{\perp\,j}\right) \D x^k, \label{eq:coeff}
\ea
where $\D x_\parallel,\,\D x_\perp^i$ are the parallel and
perpendicular components of the displacements $\D x^i$.  
Note that while we
have assumed that the ruler is small, the 
expressions for $\C,\,\B_i,\,\A_{ij}$ are valid on the 
full sky.  \reffig{SVT} illustrates the distortions induced by these
components.   
Observationally, we have 6 free parameters (assuming accurate redshifts
are available): the
location of one point $\vnhat,\,\zt$, and the separation vector described by
$\d\tilde x^i$ (with $\d\tilde x^0$ being fixed by the light cone condition).  
Using these, we can measure a (2-)scalar on the sphere, $\C$, a $2\times2$ 
symmetric matrix, $\A_{ij}$, and a 2-component vector on the sphere, $\B_i$.  
As a symmetric matrix on the sphere, $\A_{ij}$ has a scalar component, given
by the trace $\M \equiv \P^{ij} \A_{ij}$ (\emph{magnification}), and two 
components of the traceless 
part which transform as spin-2 fields on the sphere 
(\emph{shear}, ${}_{\pm 2}\g$ as defined in \refEq{shear1} below).  
These quantities are observable and gauge-invariant, although any of 
the individual contributions in \refEq{coeff} are not.  
The only exception is the proper time perturbation $\T$, 
which can be isolated by comparing two co-located rulers which evolve 
differently in time. Note that we cannot measure
any of the anti-symmetric components, such as the rotation. This is because
we have not assumed the existence of any preferred directions in the 
Universe. If there is a primary spin-1 or higher spin field, such as 
the polarization in case of the CMB, then a rotation can be measured as 
it mixes the spin$\pm 2$ components (see, e.g. \cite{gluscevic/etal:2009}).
In the next sections we study the three terms $\C,\,\B_i,\,\A_{ij}$ in turn.

\subsection{Longitudinal scalar}
\label{sec:C}

The longitudinal component can be simplified to become
\ba
\C =\:&  \frac{d\ln r_0(\tilde a)}{d\ln\tilde a} \T - \D\ln a
\left[1 - H(\zt) \frac{\partial}{\partial\zt}\left( \frac{1+\zt}{H(\zt)}\right)
\right] - A - v_\parallel + B_\parallel \vs
& 
+ \frac{1+\zt}{H(\zt)} \left(
 - \partial_\parallel A + \partial_\parallel v_\parallel 
+ \dot{B}_\parallel - \dot{v}_\parallel + \frac12 \dot{h}_{\parallel}  \right).
\label{eq:C}
\ea
The first line contains the contributions due to the fact that 
the size of the ruler evolves in time,
the scale factor at emission is perturbed from $1/(1+\zt)$, and the fact that 
the distance-redshift relation evolves, in addition to the
perturbation to the metric at the source location ($-A$) and the projection 
from coordinate-time to proper-time hypersurfaces
($B_\parallel-v_\parallel$).  The contributions from the line-of-sight
derivative of the line-of-sight displacements ($\propto (1+\zt)/H(\zt)$)
are given in the second line.  
Note the term $\partial_\parallel v_\parallel$, which is the dominant term
on small scales in the conformal-Newtonian gauge.  This term is
also responsible for the linear redshift-space distortions \cite{Kaiser87}.
Apart from the perturbation to the scale factor at emission,
$\C$ does not involve any integral terms; this is expected since $\C$
is the only term remaining if the two lines of sight coincide ($\vnhat=\vnhat'$).  
In this case, the two rays share the same path from the closer of the
two emission points, and no quantities integrated along the line of sight
can contribute to the perturbation of the ruler.  

\begin{figure}[t!]
\centering
\includegraphics[width=0.49\textwidth]{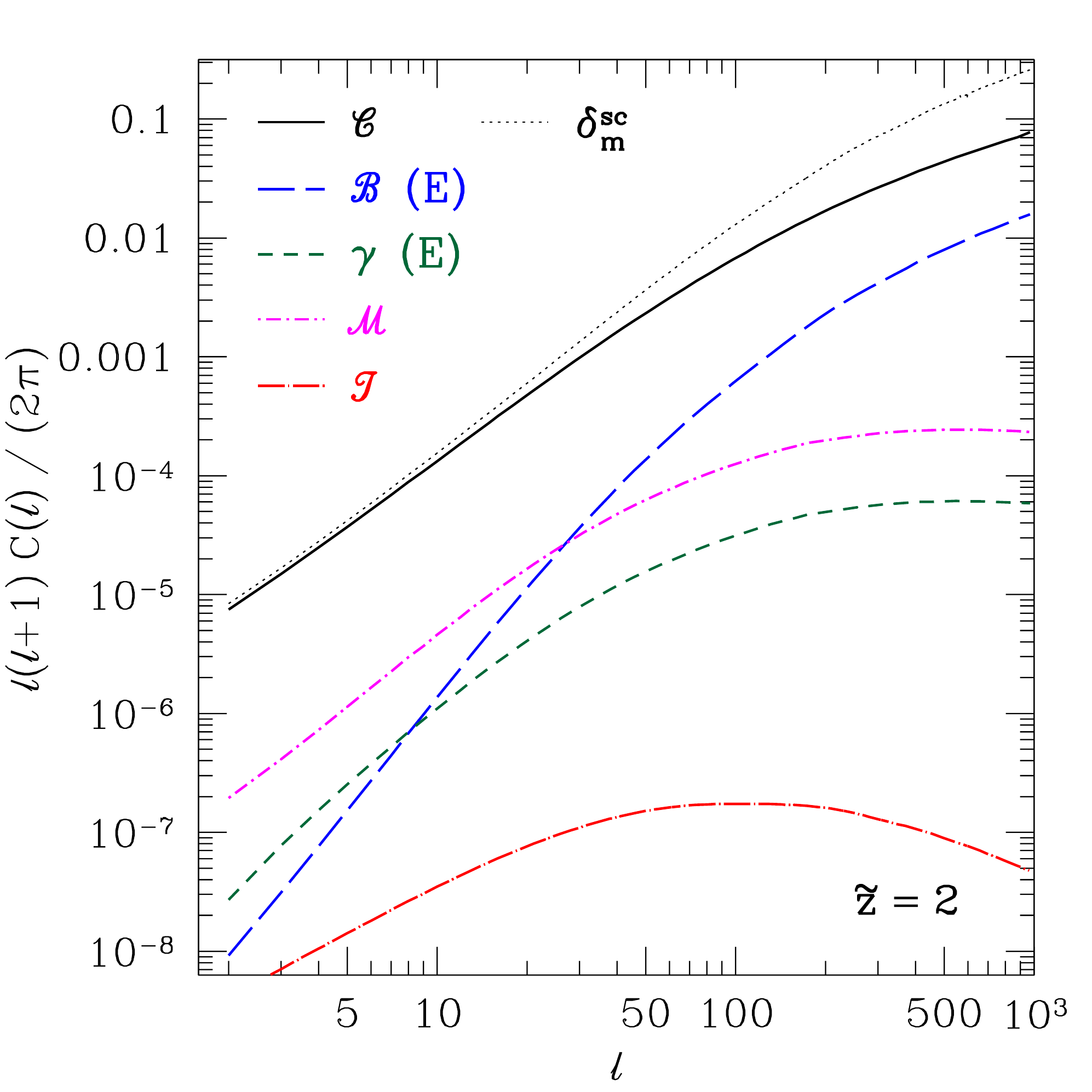}
\caption{Angular power spectra of the different standard ruler perturbations
produced by a standard scale-invariant power spectrum of curvature perturbations:
$\C$, $E$-mode of $\B_i$, $E$-mode of the shear $\gamma$, magnification $\M$, and
clock perturbation $\T$.    
$\C$ and $\M$ are calculated for a non-evolving ruler, and all are for a 
sharp source redshift of $\zt = 2$.  
For comparison, the thin dotted line shows the angular power spectrum at $z=2$ of the matter 
density field in synchronous-comoving gauge.  Note that all quantities 
shown here, except for $\d_m^{\rm sc}$, are gauge-invariant and (in principle)
observable. Adapted from \cite{stdruler,Tpaper}.
\label{fig:Cl}}
\end{figure}

Restricting to the synchronous-comoving and conformal-Newtonian gauges,
respectively, we obtain
\ba
\fl (\C)_{\rm sc} =\:&  - (\D\ln a)_{\rm sc}
\left[1 - \frac{d\ln r_0(\tilde a)}{d\ln\tilde a} - H(\zt) \frac{\partial}{\partial\zt}\left( \frac{1+\zt}{H(\zt)}\right)
\right] 
+ \frac{1+\zt}{2 H(\zt)}\dot{h}_{\parallel}. \hspace*{2cm}
\label{eq:C_sc}\\
\fl(\C)_{\rm cN} =\:& \frac{d\ln r_0(\tilde a)}{d\ln\tilde a} \T_{\rm cN} - (\D\ln a)_{\rm cN}
\left[1 - H(\zt) \frac{\partial}{\partial\zt}\left( \frac{1+\zt}{H(\zt)}\right)
\right] \vs
\fl& - \Psi - v_\parallel + \frac{1+\zt}{H(\zt)} \left(
 - \partial_\parallel\Psi + \partial_\parallel v_\parallel 
 - \dot{v}_\parallel + \dot{\Phi}  \right).
\label{eq:C_cN}
\ea
Note that in case of the sc-gauge expression, 
the redshift-space distortion term is included in the last term, through
$\dot h_\parallel/2 = \dot D + \partial_\parallel^2 \dot E$.  \reffig{Cl} shows
the angular power spectrum of $\C$ due to standard adiabatic
scalar perturbations in a $\Lambda$CDM cosmology (the details of the
calculation are given in Appendix F of \cite{stdruler}).  Clearly,
$\C$ is of the same order as the matter density contrast in 
synchronous-comoving gauge on all scales.  In particular, 
the velocity gradient term dominates over all other contributions.

\subsection{Vector}
\label{sec:vector}

Next, we have the two-component vector, which can be written as
\ba
\B_i &=\: - v_{\perp i} + B_{\perp i} + \frac{1+\zt}{H(\zt)}\partial_{\perp i}\D\ln a\,.
\label{eq:Bi}
\ea 
As expected, this vector involves the transverse derivative of the
line-of-sight displacement and the line-of-sight derivative of the
transverse displacement.  Note that these two quantities are 
\emph{not} observable individually.  

Using the spin$\pm1$ unit vectors $\v{m}_\pm$, 
$\B_i$ can be decomposed into spin$\pm1$ components:
\ba
\B_i&=\: {}_{+1}\B\: m_+^i + {}_{-1}\B\: m_-^i \vs
{}_{\pm 1}\B &\equiv\: m_\mp^i \B_i 
= - v_{\pm} + B_{\pm} + \frac{1+\zt}{H(\zt)}\partial_{\pm}\D\ln a,
\label{eq:Bpm}
\ea
where we have used the notation of \refEq{Xpm}.  
Similar to before, we can specialize this general result to the
synchronous-comoving and conformal-Newtonian gauges:
\ba
\fl
({}_{\pm 1}\B)_{\rm sc} &=\: \frac{1+\zt}{2 H(\zt)}\int_0^{\chit}d\chi\frac{\chi}{\chit}\partial_\pm \dot{h}_\parallel
\label{eq:Bpm_sc}\\
\fl
({}_{\pm 1}\B)_{\rm cN} &=\: - v_\pm + \frac{1+\zt}{H(\zt)}\partial_{\pm}\D\ln a \vs
\fl
&=\: -v_\pm + \frac{1+\zt}{H(\zt)}\Bigg(-\partial_\pm \Psi + \partial_\pm 
[v_\parallel - v_{\parallel o}] 
+ \int_0^{\chit} d\chi\frac{\chi}{\chit} \partial_\pm(\dot{\Phi}-\dot{\Psi})
\Bigg).
\label{eq:Bpm_cN}
\ea
On small scales, the dominant contribution to $\B_i$ comes from the 
transverse derivative of the line-of-sight component of the velocity
$\partial_\pm v_\parallel$, which
is of the same order as the tidal field.  

Applying the spin-lowering operator $\bar\Del$ to ${}_1\B$ yields
a spin-zero quantity (see Appendix~A of \cite{stdruler}), 
which can be expanded in terms of the usual
spherical harmonics.  We then obtain
the multipole coefficients of $\B$ as
\ba
a^{\B}_{lm}(\zt) =\:& - \sqrt{\frac{(l-1)!}{(l+1)!}} \int d\Omega\: \left[\bar\Del\,{}_1\B(\vnhat,\zt)\right] Y^*_{lm}(\vnhat).
\label{eq:aBlm}
\ea
An equivalent result is obtained for $\Del {}_{-1} \B$.  In general, the
multipole coefficients $a^{\B}_{lm}$ are complex, so that we can decompose
them into real and imaginary parts,
\be
a^{\B}_{lm} = a^{\B E}_{lm} + i\,a^{\B B}_{lm}.
\ee
One can easily show (Appendix~A of \cite{stdruler}) that under a change of parity
$a^{\B E}_{lm}$ transform as the spherical harmonic coefficients of a
vector (parity-odd), whereas $a^{\B B}_{lm}$, picking up an additional
minus sign, transform as those of a pseudo-vector (parity-even).  
These thus correspond to the polar (``$E$'') and 
axial (``$B$'') parts of the vector $\B_i$.  
As required by parity, scalar perturbations do not contribute to
the axial part $a^{\B B}_{lm}$.  
Thus, a measurement of the vector component $\B_i$ of standard ruler distortions
offers an additional possibility to probe gravitational waves with large-scale
structure, as tensor modes do contribute to $a^{\B B}_{lm}$.  
Thus, in principle the axial component of $\B_i$ could
be of similar interest for constraining
tensor modes as weak lensing $B$-modes  \cite{paperII}.  

The power spectrum of the $E$-mode of $\B$ due to standard
scalar perturbations is shown in \refFig{Cl}.  The dominant
contribution to $\B_i$ for a given Fourier mode of the matter density 
contrast in synchronous-comoving gauge is $\propto k_\perp k_\parallel/k^2\, \d^{\rm sc}_m(\vk,\zt)$, while the corresponding contribution to the longitudinal
scalar $\C$ is $\propto k_\parallel^2/k^2\, \d^{\rm sc}_m(\vk,\zt)$.  
Even though approximate scaling arguments suggest that 
$C_\C(l)$, $C^{EE}_{\B}(l)$ should scale roughly equally with $l$, the 
different structure in terms of $k_\parallel,\,k_\perp$ (together with the
shape of the matter power spectrum in $\Lambda$CDM) leads to 
a faster scaling of $C_{\B}(l)$ with $l$ for $l \lesssim 500$
(see discussion in \cite{stdruler}).  

\subsection{Transverse tensor: shear and magnification}

Finally, we have the purely transverse component,
\ba
\fl
\A_{ij} =\:& \frac{d\ln r_0(\tilde a)}{d\ln\tilde a} \T\,\P_{ij} - \D\ln a \: \P_{ij}
- \frac12 \P_i^{\  k} \P_j^{\  l} h_{kl} 
%
- \partial_{\perp\,(i} \D x_{\perp\,j)}
- \frac{1}{\chit} \D x_\parallel \P_{ij},\hspace*{1cm}
\label{eq:Aij}
\ea
where we have again inserted projection operators for clarity (note that
$\P_{ij}$ serves as the identity matrix on the sphere).  As
a symmetric matrix on the sphere, $\A_{ij}$ has a scalar component, given
by the trace $\A$, and two components of the traceless part which transform
as spin-2 fields on the sphere.  The trace corresponds to the change in
area on the sky subtended by two perpendicular standard rulers.  Thus,
it is equal to the magnification $\M$ (see also \reffig{SVT}).  
The two components of the traceless
part correspond to the shear $\g$.  If we choose a fixed coordinate
system $(\v{e}_\theta,\v{e}_\phi,\vnhat)$, we can thus write
\be
\A_{ij} = \left(\begin{array}{cc}
\M/2 + \g_1 & \g_2 \\
\g_2 & \M/2 - \g_1
\end{array}\right).
\label{eq:Aijcoord}
\ee
Below, we will derive the magnification and shear for the general 
perturbed FRW metric \refEq{metric}.

\subsubsection{Magnification}
\label{sec:mag}

Taking the trace of \refEq{Aij} yields
\ba
\fl
\M \equiv\:& \P^{ij} \mathcal{A}_{ij} 
%
= 2 \frac{d\ln r_0(\tilde a)}{d\ln\tilde a} \T - 2\D\ln a 
- \frac12 \left(h - h_\parallel\right) 
- \frac{2}{\chit} \D x_\parallel + 2\hat\k\,.
\hspace*{1cm}
\label{eq:mag}
\ea
The magnification is directly related to the fractional perturbation in
the angular diameter and luminosity
distances (see \cite{HuiGreene,BonvinEtal06}) through 
\be
\frac{\Delta D_L}{D_L} = \frac{\Delta D_A}{D_A} = -\frac12 \M,
\ee
where the first equality for the luminosity distance follows from 
the conservation of the photon phase space density.  That is, $\M$ describes
both the change in apparent angular size of a spatial ruler as well as the 
change in observed flux of a standard candle.  
The contributions to the magnification are straightforwardly interpreted
as coming from the time evolution of ruler scale (or intrinsic source luminosity, if applied to standard candles) through the proper time perturbation $\T$; 
from the conversion of coordinate distance to physical scale at
the source (both from the perturbation to the scale factor $\D\ln a$ and the
metric at the source projected perpendicular to the line of sight, 
$h-h_\parallel$);  from the fact that the entire ruler is moved
closer or further away by $\D x_\parallel$;  and finally from the coordinate
convergence $\hat\k$ defined through
\be
\hat\k = -\frac12 \partial_{\perp\,i} \D x_\perp^i.
\label{eq:kappadef}
\ee
This term is the dominant contribution to $\M$ on small scales.  However,
the coordinate convergence is a gauge-dependent quantity; 
see for example Appendix~B2  in \cite{gaugePk}.  
In conformal-Newtonian gauge, it assumes its familiar form,
\ba
(\hat\k)_{\rm cN} =\:& - v_{\parallel o}
+ \frac12 \int_0^{\chit} d\chi\,\frac{\chi}{\chit} 
 (\chit-\chi)
\nabla_\perp^2 \left(\Psi - \Phi\right)\,,
\label{eq:kcN}
\ea
with an additional term $-v_{\parallel o}$ contributing to the dipole
of $\hat\k$ only, which corresponds to the relativistic aberration effect
at linear order.  An explicit expression for the magnification 
in general gauge is straightforward to obtain, however it becomes lengthy.  
Here we just give the results for the two most popular gauge choices.
First, in synchronous-comoving gauge [\refEq{metric_sc}] we obtain  
\be
(\M)_{\rm sc} = 2 \left[\frac{d\ln r_0(\tilde a)}{d\ln\tilde a} - 1\right](\D\ln a)_{\rm sc} 
-\frac12 (h - h_\parallel)
+ 2 (\hat\k)_{\rm sc} - \frac2{\chit} \D x_\parallel\,,
\label{eq:magsc}
\ee
where
\ba
\fl
(\hat\k)_{\rm sc} = -\frac14\left[h_o - 3 (h_\parallel)_o \right]  \label{eq:kappasc}
+\frac12 \int_0^{\chit} d\chi\,\Bigg[
 (\partial_\perp^l h_{lk}) \nhat^k 
+ \frac{1}{\chi} \left(h - 3 h_\parallel\right)
 - \frac12 (\chit-\chi)\frac{\chi}{\chit}  \nabla_\perp^2 h_\parallel
\Bigg]. \nonumber
\ea
In conformal-Newtonian gauge [\refEq{metric_cN}], we have $(h - h_\parallel)/2 = 2\Phi$,
so that the magnification in this gauge becomes
\ba
\left(\M\right)_{\rm cN} =\:& 2 \frac{d\ln r_0(\tilde a)}{d\ln\tilde a} \T_{\rm cN} + \left[- 2 + \frac2{a H \chit}\right] (\D\ln a)_{\rm cN} - 2 \Phi  
+ 2(\hat\k)_{\rm cN} \vs
& - \frac{2}{\chit} \int_0^{\chit} d\chi\: (\Psi-\Phi)
+ \frac2{\chit} \int_0^{t_o} dt\: \Psi(\v{0},t)\,.
\label{eq:magcN}
\ea
The last term in \refEq{magcN} is a pure monopole and thus usually absorbed 
in the ruler calibration 
(since $r_0$ can rarely be predicted from first principles without
any dependence on the background cosmology).  Nevertheless, the monopole of $\M$
is in principle observable, and including this term ensures that gauge modes 
(for example superhorizon metric perturbations) do not affect its value.  

We have thus arrived at a general gauge-invariant expression for the 
magnification without having to perform lengthy calculations.  Moreover,
the physical starting point from a standard ruler scale has allowed us
to identify a previously overlooked contribution to the magnification.  
This contribution is given by the clock variable $\T$ and becomes
relevant whenever the ruler scale evolves over cosmic time.  In many
applications, this is the case, although $\T$ is sub-dominant to
the lensing convergence $\hat\k$ on all but the largest scales \cite{Tpaper}.  

\subsubsection{Shear}
\label{sec:shear}

We now consider the traceless part of $\A_{ij}$, given by
\ba
\g_{ij}(\vnhat) &\equiv\: \A_{ij} - \frac12\P_{ij} \M \vs
&=\: 
-\frac12 \left(\P_i^{\  k} \P_j^{\  l} 
- \frac12 \P_{ij} \P^{kl} \right) h_{kl} 
- \partial_{\perp (i} \D x_{\perp\,j)} 
- \P_{ij} \hat\k.
\label{eq:shear1}
\ea
Here, the terms $\propto \P_{ij}$ in \refEq{Aij} drop out.  
The last two terms here are what commonly is regarded as the shear, 
i.e. the trace-free part of the transverse derivatives of the transverse
displacements. The first term on the other hand is important to 
construct an \emph{observable} as it ensures a gauge-invariant result.  
This is the term referred to as
``metric shear'' in \cite{DodelsonEtal}.  Its physical significance
becomes clear when constructing the Fermi normal coordinates, or local inertial
frame, for the region containing the standard ruler.  

Consider a region of spatial extent $R$, say centered on a given galaxy, 
and enclosing our standard ruler.  
We can construct orthonormal Fermi normal coordinates \cite{Fermi,ManasseMisner}
around the center of this region, which follows a timelike geodesic,
by choosing the origin to be located at the center of the region at all times,
and the time coordinate to be the proper time along the geodesic.  
As a local inertial frame, the spacetime in the Fermi coordinates 
$(t_F, x^i_F)$ is Minkowski close to the geodesic, 
with corrections proportional to $\vx_F^2/R_c^2$ 
where $R_c$ is the curvature scale of the spacetime.  
Thus, as long as these corrections to the metric are negligible, 
there is no preferred direction in this frame, and the size $r_0$ of the standard ruler 
has to be (statistically) independent of the orientation. 
The most obvious example is galaxy shapes, which are used for cosmic shear
measurements. In the Fermi frame, when neglecting tidal alignments,
galaxy orientations are statistically random.  
As shown in \cite{stdruler,paperII}, the transformation from global coordinates
to Fermi coordinates for a purely spatial metric perturbation $h_{ij}$ 
is given by
\be
a^{-1} x_{F}^i 
= x^i + \frac12 h_{ij}(0) x^j + \O(\partial_m h_{kl} x^2).
\label{eq:transf}
\ee
In order to obtain the shear relative to the Fermi frame, we need to add the
transformation \refEq{transf} to the displacements $\D x^i$:
\be
\D x^i \rightarrow \D x^i + \frac12 h_{ij}(0) x^j \,.
\ee
With these new displacements, the transverse derivative of the
transverse displacement becomes
\ba
\partial_{\perp (i} \D x_{\perp\,j)} \rightarrow \partial_{\perp (i} \D x_{\perp\,j)} + \frac12 \P_i^{\  k} \P_j^{\  k} h_{kl} + \O(\partial_k h_{ij} [\tilde\vx-\tilde\vx'])\,,
\label{eq:shearF}
\ea
where the last term is negligible in the small-ruler approximation.  
This agrees exactly
with the result derived above, \refEq{shear1} [after subtracting the
trace of \refEq{shearF}].  
Note that the Fermi coordinates are uniquely
determined up to three Euler angles.  The statement that galaxy orientations
are random in this frame is thus coordinate-invariant.  

$\g_{ij}$ is a symmetric trace-free tensor on the sphere, and can thus
be decomposed into spin$\pm 2$ components (in analogy to the 
polarization of the CMB).  Following Appendix~A in \cite{stdruler} 
(see also \cite{Hu2000})
we can write $\g_{ij}$ as
\ba
\g_{ij} &=\: {}_2\g\, m_+^i m_+^j + {}_{-2}\g\, m_-^i m_-^j \vs
{}_{\pm 2}\g &=\:  m_\mp^i m_\mp^j \g_{ij}\,,
\label{eq:sheardecomp}
\ea
where ${}_{\pm2}\g$ are spin$\pm2$ functions on the sphere (in analogy
to the combination of Stokes parameters $Q \pm i U$).  The general,
lengthy expression for the shear components can be found in \cite{stdruler}.  
Here we give the expressions for the synchronous-comoving (sc) and 
the conformal-Newtonian (cN) gauges (note that $h_\pm = 0$ in cN gauge):
\ba
\fl
\left({}_{\pm 2}\g\right)_{\rm sc} =\:&
-\frac12 h_\pm - \frac12  (h_{\pm})_o - \int_0^{\chit} d\chi\, \Bigg[
\left(1-2\frac{\chi}{\chit}\right) (\partial_\pm h_{kl}) m_\mp^k \nhat^l
- \frac{1}{\chit} h_\pm
\label{eq:shear_sc}
\\
\fl
& \hspace*{4.4cm} 
+ (\chit-\chi)\frac{\chi}{\chit}
\frac12 (m_\mp^i m_\mp^j \partial_i\partial_j h_{lk})\nhat^l\nhat^k
\Bigg]
\hspace*{0.5cm} \vs
\fl
\left({}_{\pm 2} \g\right)_{\rm cN} =\:& 
\int_0^{\chit} d\chi\, (\chit-\chi)\frac{\chi}{\chit}
 m_\mp^i m_\mp^j \partial_i\partial_j 
\left(\Psi - \Phi \right)
\,.
\label{eq:shear_cN}
\ea
We see that \refEq{shear_cN} recovers the ``standard'' result; in other words, there are no additional
relativistic corrections to the shear \emph{in cN gauge}.  This is not
surprising following our arguments above: in the conformal-Newtonian gauge,
the transformation \refEq{transf} from global coordinates to the local Fermi
frame is isotropic since $h_{ij} = 2\Phi \d_{ij}$.  Thus, it does not
contribute to the shear.  Note however that this only applies to
scalar perturbations;  when
considering vector or tensor perturbations, \refEq{shear_sc} is the
relevant expression which does contain terms beyond the derivative
of the deflection angle.  This is also of relevance to studies of
gravitational lensing of the CMB by gravitational waves \cite{Book/etal}
(see also \cite{paperII}).  

\reffig{Cl} shows the angular power spectrum of shear and magnification
due to scalar perturbations
for a sharp source redshift $\zt = 2$.  For $l\gtrsim 10$, the results
follow the familiar relation $C_{\M}(l) = 4 C^{EE}_\g(l)$, valid when
all relativistic corrections to the magnification become irrelevant
so that $\M \simeq 2\hat\k$.  
These corrections slightly increase the magnification for small $l$.  
We also see that $\g$ and $\M$ are suppressed with respect to $\C$
and $\B$ (on smaller scales), at least when the latter are evaluated
for a sharp source redshift.  This is a well-known consequence of
the projection with the broad lensing kernel, leading to a cancellation
of modes that are not purely transverse (see e.g. \cite{JeongSchmidtSefusatti}).  

\subsection{Applications}
\label{sec:app}

Consider a galaxy whose image projected on the sky, as seen by a local
observer, has an ``intrinsic'' intensity or surface brightness $I(\vtheta)$
(here $\vtheta=0$ corresponds to the centroid of the galaxy).  
Then, the ruler formalism immediately yields the observed intensity through
\be
I_{\rm obs}(\tilde{\vtheta}^i) = I\left(\tilde{\vtheta}^i - \A^i_{\  j} \tilde{\vtheta}^j\right) = \left[1 - \A_i^{\  j} \tilde{\vtheta}^i \frac{\partial}{\partial\tilde{\vtheta}^j} \right] I(\tilde{\vtheta})
+ \O([\A_{ij}]^2)\,,
\ee
where $\A_{ij}$ is the sky-plane projection of the ruler perturbations, \refEq{coeff}, and we have expanded to linear order.  This is the well known effect
of weak gravitational lensing on an image.  

Now consider the case of a spectroscopic survey, where we measure the
small-scale correlation function $\tilde\xi(\tilde{\vr}, \zt)$ as a function
of the three-dimensional separation vector $\tilde{\vr}$ and the redshift $\zt$.  
As shown in \cite{pajer/schmidt/zaldarriaga}, the observed correlation function is
given in terms of the expectation value of the intrinsic correlation function $\xi(r, z)$ by
\ba
 \tilde\xi(\tilde \vr, \tilde\tau) 
=\:& \left[1 - a_{ij}(\tilde x) \tilde\vr^i \partial_{\tilde r}^j + \T(\tilde x)\partial_{\zt}  + 2 \<\tilde\d\>(\tilde x)\right] \xi(\tilde\vr;\tilde z)\,,
\ea
where $\<\tilde\d\>$ is the mean observed overdensity of the tracer within the volume over which
$\tilde\xi$ is measured (see the next section), which simply serves to
rescale the local mean density.  The tensor $a_{ij}$ contains the ruler perturbations:
\ba
a_{ij} =\:& \C\, \nhat_i \nhat_j + \nhat_{(i} \P_{j)k} \,\B^k + \P_{ik} \P_{jl} \,\A^{kl}\,,\label{eq:rulerpert}
\ea
where we have inserted trivial projection operators for $\B,\,\A$.  

These examples serve to illustrate how the standard ruler formalism
can be immediately applied to predict cosmological observables.

\section{Galaxy clustering}
\label{sec:clustering}

The statistics (correlation functions) of large-scale structure tracers have a long history 
as one of the most important observational tools in cosmology.  
The fundamental
building block of these statistics is the observed number density $\tilde n_g$
of tracers inferred from their apparent positions on the sky and redshifts.  
In this section, we show how $\tilde n_g$ is related to the spacetime
perturbations in the relativistic context.  For this, we need to 
consider two effects:  first, the effect of spacetime perturbations on the 
propagation of light emitted from the sources systematically distorts the 
observed galaxy density contrast \cite{yoo/etal:2009,yoo:2010,
bonvin/durrer:2011,challinor/lewis:2011,BaldaufEtal,gaugePk,paperI}.
Second, we need to relate the number density of galaxies to the matter
density, a procedure commonly knows as \emph{biasing}, which involves
additional subtleties in the relativistic context \cite{BaldaufEtal,gaugePk}.  

As discussed earlier in \refsec{deflection}, observers chart galaxies 
according to the observed position $\tilde{x}^\mu=(\eta_0-\chit,\vnhat\chit)$.
The galaxy density is estimated based on the observed spatial coordinate 
$\tilde{\vx}=\chit\vnhat$, and then compared with the \emph{mean} number density
$\bar{n}_g(\zt)$ at fixed observed coordinate to infer the
local galaxy overdensity $\tilde\d_g$. 
Once corrected for window function and selection effects, the mean galaxy 
number density only depends on the observed redshift.  
In this sense, the observed density contrast $\tilde\d_g$ can be seen as
defined in a constant-observed-redshift gauge.  
Throughout, we will assume that $\bar{n}_g(z)$ corresponds to the true
mean density of galaxies, i.e. we will neglect the effect of super-survey
modes.  

The number of galaxies enclosed in a spatial volume $V$ defined in the
observed coordinates is given by
\be
N(V) = \int_V d^3\tilde{\vx}\; 
\sqrt{-g(x)} n_g(x) \varepsilon_{\mu\nu\rho\sigma}u^\mu(x)
\frac{\partial x^\nu}{\partial\tilde{x}^1}
\frac{\partial x^\rho}{\partial\tilde{x}^2}
\frac{\partial x^\sigma}{\partial\tilde{x}^3}\,,
\label{eq:N1}
\ee
where we have transformed the integral to observed coordinates $\tilde{x}$,
$V$ is a spatial volume on a constant-observed-redshift slice,
and $x(\tilde x)$ denotes the true spacetime location corresponding to
the observed location $\tilde x$.  $n_g$ is the physical (as opposed to
comoving) number density of tracers in the perturbed FRW coordinates
[\refeq{metric}].  

We now employ a useful trick.  Rather than expressiong $n_g$ in terms of
the galaxy density perturbation $\d_g$ in some arbitrary gauge, we 
fix the coordinates to the \emph{constant-observed-redshift} (``or'') \emph{gauge}.  
We thus write $n_g$ in term of the mean number density $\bar{n}_g$ and the
perturbation $\d^{\rm or}_g$ to the comoving number density in 
the constant-observed-redshift gauge as
\be
a^3 n_g(x) = \tilde{a}^3
\bar{n}_g(\zt)
\left[
1 + \delta^{\rm or}_g(\vx, \zt)
\right]\,,
\label{eq:deltag1}
\ee
where $\zt$ is the observed redshift corresponding to the spacetime
location $x$, and $\tilde a = 1/(1+\zt)$.  Since $\d_g^{\rm or}$ is already 
first order, we can neglect the distinction between $\vx(\tilde x)$ and
$\tilde\vx$ in its argument.  \refeq{deltag1} can be understood as the
definition of $\d^{\rm or}_g$.  

We rewrite the right hand side of \refeq{N1} in terms of the metric 
perturbation to linear order as
\ba
\fl
N(V)
&= \int_V d^3\tilde{\vx}\:
\left(1+A+\frac{h}2\right)
\tilde{a}^3\bar{n}_g(\zt)
\left[1+\delta^{\rm or}_g(\tilde{\vx},\zt) \right]
\left(
(1-A)\left|\frac{\partial x^i}{\partial \tilde{x}^j}\right|
+
v_\parallel
\right).
\ea
The observed galaxy number density $\tilde n_g$, on the other hand, satisfies
by definition
\be
N = \int d^3 \tilde{\vx}\: \tilde{a}^3\tilde{n}_g(\tilde{\vx},\zt)\,.
\ee
By equating the two, we find the observed galaxy density contrast as
\be
\fl\tilde{\delta}_g(\tilde{\bf x})
\equiv
\frac{\tilde{n}_g(\tilde{\bf x},\tilde{z})}{\tilde{n}_g(\zt)}
-1
=
\delta^{\rm or}_g(\tilde{\vx},\zt) 
+\frac{h}2
+ \partial_{\parallel}\Delta x_\parallel
+
\frac{2\Delta x_\parallel}{\tilde{\chi}} - 2 \hat{\kappa}
+
v_\parallel\,.
\label{eq:deltag2}
\ee
Here, we have used the Jacobian 
\be
\left|\frac{\partial x^i}{\partial \tilde{x}^j}\right|
=
1+\frac{\partial \Delta x^i}{\partial \tilde{x}^i}
=
1 
+ \partial_{\parallel}\Delta x_\parallel
+
\frac{2\Delta x_\parallel}{\tilde{\chi}} - 2 \hat{\kappa}
\ee
with the lensing convergence $\hat{\kappa}$ defined in \refEq{kappadef}.  
All contributions in \refeq{deltag2} apart from $\d_g^{\rm or}$ thus correspond
to the apparent modulation of the galaxy abundance due to volume distortion
effects.  

Next, we have to relate $\delta^{\rm or}_g$ in \refEq{deltag2} to the matter
density through a biasing relation.  
The galaxy density contrast $\delta^{\rm or}_g$ in \refEq{deltag2} is defined in the 
constant-observed-redshift slicing, while the linear bias relation
between galaxy density contrast and matter density contrast holds 
only on \emph{constant-proper-time} (``pt'') slices.  This is because in the
large-scale limit, galaxies only know about the local age of the Universe
and the local matter density 
(see \cite{BaldaufEtal} and \S III in \cite{gaugePk} for a detailed discussion).\footnote{This assumes that there are no additional degrees of freedom relevant on large scales, such as dark energy perturbations, neutrinos, or fifth forces.  The impact of these on the general linear biasing relation is an interesting question, though beyond the scope of this review.} The shift between
the constant-observed-redshift slice and constant-proper-time slice is
given by the observable $\T$ that we have discussed in \refsec{clock}.  
Then, the relation between $\delta^{\rm or}_g(\tilde{\vx},\zt)$ and the matter
perturbation $\d_m^{\rm pt}$ in the constant-proper-time (or synchronous) gauge is given
by the standard linear gauge transformation
\be
\delta^{\rm or}_g(\tilde{\vx},\zt) 
= b\,\delta_m^{\rm pt} 
 + \frac{d(a^3 \bar{n}_g)}{d\ln a} \T
\equiv b\,\delta_m^{\rm pt} 
 + b_e \T,
\label{eq:biasing}
\ee
where we have introduced the dimensionless parameter $b_e$
quantifying the evolution of the mean comoving number density of tracers.  
Note that this relation only involves observable quantities, so that
both $b$ and $b_e$ are well defined and gauge-invariant.  It also serves as the
unambiguous starting point for extending the bias relation to higher order
in perturbations, for example by adding a term $b_2/2\, (\d_m^{\rm pt})^2$ to
the right hand side.  
 
Finally, $\d_m^{\rm pt}$ is related to the matter density
perturbation $\d_m$ in the chosen gauge through
\be
\delta_m^{\rm pt} 
= \delta_m + 3 \tilde H\int_0^{\tilde{\eta}} A(\vx,\eta)a(\eta) d\eta\,.
\ee
Combining the last two equations, we find the galaxy density contrast on the 
constant-observed-redshift slice in terms of the density contrast in 
an arbitrary gauge as
\be
\delta^{\rm or}_g(\tilde{\vx},\zt) 
=
b\left[\delta_m 
+ 3\tilde{H}\int_0^{\tilde{\eta}} A(\vx,\eta)a(\eta) d\eta\right]
+ 
b_e \T.
\ee
This yields our final expression: 
\be
\fl
\tilde \d_g(\tilde{\vx},\tilde z) 
= b\left[\delta_m + 3\tilde{H}\int_0^{\tilde{\eta}} A(\vx,\eta)a(\eta) d\eta\right]
+ b_e \,\T 
+ \frac12 h + \partial_{\chit} \D x_\parallel + \frac{2\D x_\parallel}{\chit} - 2\hat\k + v_\parallel\,.
\label{eq:dgtilde}
\ee
Here,
\ba
\partial_{\chit} \D x_\parallel 
=\:&
A - B_\parallel - \frac12 h_\parallel
- H(\zt)\frac{\partial}{\partial\zt}\left(\frac{1+\zt}{H(\zt)}\right) \Delta\ln a \vs
&
- \frac{1+\zt}{H(\zt)}
\left(
-\partial_\parallel A
+
\partial_\parallel v_\parallel
-\dot{v}_\parallel
+ \frac12 \dot{h}_\parallel + \dot{B}_\parallel
\right)\,.
\ea

One subtlety we have neglected so far is that observational selection 
effects can modify the observed galaxy density, \refEq{dgtilde}.  
Usually surveys observe galaxies above a certain magnitude threshold.
Weak lensing magnifies/de-magnifies the flux of the source galaxies and 
therefore induces another contribution to the observed galaxy density
(\emph{magnification bias}).  
For a population of galaxies at fixed redshift $\zt$ with cumulative 
luminosity function $\bar{n}(>L_{\rm min})$, we define 
\be
\Q \equiv -\frac{d\ln \bar{n}(>L_{\rm min})}{d\ln L_{\rm min}} \,,
\ee
where $\M$ is the magnification discussed in detail in \refsec{mag}.  
Then, the contribution to $\tilde\d_g$ induced by the lensing magnification [\refEq{mag}] is $\Q\,\M$.  Note that our ``ruler'' here is the luminosity of
galaxies at the cutoff $L_{\rm min}$, so that $d\ln r_0/d\ln a$ in \refeq{mag} is to be
replaced with the evolution of the intrinsic luminosity of galaxies with
$L = L_{\rm min}$ at $\zt$, $d\ln L_{\rm min}/d\ln a$, in order to take the 
evolving ruler effect into account (\refsec{evolving_ruler}).  
We finally obtain the observed density contrast including magnification bias as
\ba
\fl
\tilde \d_g(\tilde{\vx},\tilde z) 
=& b\left[\delta_m + 3\tilde{H}\int_0^{\tilde{\eta}} A(\vx,\eta)a(\eta) d\eta\right]
+ \left(b_e + 2\Q\left[\frac{d\ln L_{\rm min}}{d\ln a}-1\right]\right) \,\T 
\vs
\fl
&+ 2\Q\tilde{H}\int_0^{\tilde{\eta}} A(\vx,\eta)a(\eta) d\eta
+ \frac12(1-\Q) h + \frac{\Q}{2}h_\parallel + \partial_{\chit} \D x_\parallel 
+ (1-\Q)\frac{2}{\chit} \D x_\parallel \vs
\fl
& - 2(1-\Q)\hat\k + v_\parallel\,.
\label{eq:dgtilde_M}
\ea

\refeq{dgtilde_M} provides the complete result for the observed overdensity
of a tracer at linear order in a general gauge; to our knowledge, 
this is the first time an expression for $\tilde\d_g$
has been given in a general gauge with a physical treatment of galaxy bias.  
When restricted to conformal-Newtonian gauge, this agrees with
\cite{challinor/lewis:2011,BaldaufEtal} (note the discussion around Eq.~(31)
of the former reference);  restricting to synchronous-comoving gauge 
yields the results derived in \cite{gaugePk}.  
Note also that all previous references implicitly assumed that 
$d\ln L_{\rm min}/d\ln a$, which induces an
apparent density contrast due to time evolution of the luminosity function, 
is zero (though since this term is proportional to $\T$, it is expected
to be subdominant on all scales, see \reffig{Cl}).  
Throughout, we have assumed a sharp source redshift.    
The projection over a wider photometric redshift bin is straightforward.  

Assuming the coefficients $b,\,b_e,\,\Q$ are all of order unity, the
various terms in \refeq{dgtilde_M} can be ranked in terms of relative
importance according to their scaling in Fourier space.  The largest
terms, ``order 1'', are
\be
\tilde\d_g^{\O(1)} = b\, \d_m + \frac{1+\zt}{\tilde H} \partial_\parallel v_\parallel - 2(1-\Q) \hat\k\,.
\label{eq:dgleading}
\ee
Here, we have assumed that one of the standard gauges is chosen where
$\d_m \simeq \d_m^{\rm pt}$ on small scales (this includes conformal-Newtonian gauge).  
\refEq{dgleading} is the standard small-scale result for the apparent galaxy overdensity,
including the leading redshift-space distortion (``Kaiser formula'' \cite{Kaiser87}) and magnification bias.  Next, there are contributions suppressed
by $\tilde a \tilde H/k$ (``velocity-type'') and $(\tilde a \tilde H/k)^2$
(``potential-type'').  The potential-type contributions have the same
$k$-dependence as the scale-dependent bias from primordial non-Gaussianity
of the local type \cite{DalalEtal08}.  These are numerically the smallest
contributions, and amount to the effect of local primordial non-Gaussianity
with $f_{\rm NL} \lesssim 1$, as shown in \cite{gaugePk}.  The velocity-type
contributions, which are $\propto v_\parallel$ and $\partial_\parallel \Psi$
in case of conformal-Newtonian gauge, 
are larger and likely to be measurable in upcoming surveys \cite{Hamaus/etal:12}.  

\section{Conclusion and future work}\label{sec:conclusion}

In this paper, we have derived the effects of light deflection in the perturbed
FRW universe, and an associated set of observables in the large-scale 
structure of the Universe.  Light deflection distorts the observed position and 
redshift of cosmic events, and
such distortions can be measured for events with known cosmic age 
(cosmic clock) or length scale (cosmic ruler). Distortions in the cosmic
clocks are described by the observable $\T$, which is the redshift perturbation between
the constant-proper-time slicing and the constant-observed-redshift slicing
(an example being the CMB temperature perturbations in the Sachs-Wolfe limit);
distortions in the cosmic rulers are completely described by six observables which are classified as
two scalars ($\C$ and the magnification $\M$), two components of a divergence-free vector 
($\B_i$), and two components of a tensor (shear $\gamma_{ij}$) 
on the celestial sphere.

We have also presented the fully general expression for the observed galaxy density 
contrast at linear order, a fundamental galaxy clustering observable, including the 
volume distortion due to the light deflection, evolving number density,
galaxy density bias, as well as the magnification bias generalized to evolving 
luminosity function.

All expressions in this paper are derived at linear order in 
density, velocity, and metric perturbations, but in their
most general form, sometimes referred to as \emph{gauge ready} form.  
Therefore, all expressions in this paper can be trivially restricted to any 
specific gauge.  We show gauge-fixed examples for the 
conformal Newtonian (cN) gauge and synchronous comoving (sc) gauge in 
\refsec{ruler}.  
Extending the calculations to higher order should also be
straightforward, albeit tedious, by following the logical steps described in this paper.  
In fact, three pre-prints (\cite{2ndorder1,2ndorder2,2ndorder3}) extending the calculation of
the observed galaxy density contrast to second order have appeared 
by the time this paper was written. 

In summary, we now have a general relativistic description for the complete set
of large-scale observables of the large-scale structure
($\T$, $\C$, $\B_i$, $\M$, $\gamma_{ij}$ as well as $\tilde\delta_g$).  Hence,
future work can focus on the applications of these results.  
In particular, in conjunction with future large-scale structure surveys 
mapping a significant fraction of the observable universe ($V\gtrsim 100~[{\rm Gpc}/h]^3$) 
such as Euclid, LSST and SKA, we envisage three directions that this 
line of research should pursue:

First of all, all expressions shown in this paper, with the exception of the
biasing relation, \refeq{biasing}, only depend on kinematics of
light propagation, and, therefore, hold for \emph{any} metric theory 
of gravity.  Non-smooth Dark Energy as well as modified gravity theories 
predict different relations between
the aforementioned observables and the cosmic density perturbations than the 
standard $\Lambda$CDM or smooth Dark Energy scenarios.  
This effect has only recently been explored for galaxy clustering \cite{lombriser/etal:2013}, and it would be interesting
to see how large the impact could be in other large-scale structure observables.

The relativistic effects discussed here appear on near horizon scales 
$k\sim 0.001~{h/{\rm Mpc}}$.  Our expressions for 
$\T,\,C,\,\B,\,\M,\,\g,\,\tilde\d_g$ are valid on the full sky 
and can immediately be used to calculate angular auto- and cross-correlations
$C(\l)$ on arbitrarily large scales.  However, for spectroscopic surveys, angular correlations in 
many narrow redshift bins might not be the optimal approach.  While
on sufficiently small scales the flat-sky approximation can be used,
on large scales conventional Fourier-basis decompositions must be modified 
to include the 
effects from sky curvature as well as time evolution of the cosmic structure
\cite{Yoo/Desjacques:2013,DiDio/etal:2013,DiDio/etal:2014}.

Finally, going beyond the power spectrum, it is interesting to investigate
the impact of relativistic effects on higher order correlation functions.  
In general, this requires the calculation of the relativistic
observables to higher order, e.g. second order for the  bispectrum of cosmic 
shear and galaxy clustering.  However, most of the signal-to-noise in the
bispectrum on very large scales is in the squeezed limit, where one large-scale
mode is correlated with two small-scale modes.  In this limit, one can
use a trick to circumvent the second order calculation \cite{pajer/schmidt/zaldarriaga}.  
The bispectrum in this limit is then entirely determined by the 
linear order ruler distortions of the small-scale correlation function
that are described in \refsec{app} (along with any primordial contribution
due to local-type primordial non-Gaussianity).  

In summary, the relativistic effects that we discuss here must be included 
whenever very large scale modes are measured, and are thus crucial in order
to fully exploit the information in future large-scale structure surveys.  
Fortunately, they can be accurately predicted in terms of only a few
tracer-dependent free parameters.


\section*{References}
\bibliography{GW}

\end{document}